\documentclass[aps,prl,onecolumn,superscriptaddress,notitlepage,groupedaddress]{revtex4-1}

\usepackage{graphicx}
\usepackage{mathptmx, textcomp}
\usepackage[usenames]{xcolor}
\usepackage{dcolumn}
\usepackage{bm}
\usepackage{amssymb, amsmath}
\usepackage[colorlinks,urlcolor=blue,citecolor=blue,linkcolor=blue]{hyperref}
\usepackage{cleveref}
\usepackage{siunitx}
\usepackage{physics}

\begin{document}

\title{Dilute Fluid Governed by Quantum Fluctuations}

\author{Nils B. J\o rgensen}
\author{Georg M. Bruun}
\author{Jan J. Arlt}

\affiliation{Institut for Fysik og Astronomi, Aarhus Universitet, 8000 Aarhus C, Denmark.}
\date{\today}

\begin{abstract}
Understanding the effects of interactions in complex quantum systems beyond the mean-field paradigm constitutes a fundamental problem in physics. Here, we show how the atom numbers and interactions in a Bose-Bose mixture can be tuned to cancel mean-field interactions completely. The resulting system is entirely governed by quantum fluctuations -- specifically the Lee-Huang-Yang correlations. We derive an effective one-component Gross-Pitaevskii equation for this system, which is shown to be very accurate by comparison with a full two-component description. This allows us to show how the Lee-Huang-Yang correlation energy can be accurately measured using two powerful probes of atomic gases: collective excitations and radio-frequency spectroscopy. Importantly, the behavior of the system is robust against deviations from the atom number and interaction criteria for canceling the mean-field interactions. This shows that it is feasible to realize a setting where quantum fluctuations are not masked by mean-field forces, allowing investigations of the Lee-Huang-Yang correction at unprecedented precision.
\end{abstract}

\maketitle

\twocolumngrid

A key challenge in modern physics is to understand the properties of interacting quantum systems. Except in the limit of weak interactions, where a mean-field approach is often sufficient, theories for interacting systems typically involve approximations or brute force numerical calculations. The Lee-Huang-Yang (LHY) correction to the ground-state energy of a Bose gas~\cite{lee1957} is a seminal result, which goes beyond mean-field theory by including quantum fluctuations. In spite of its fundamental importance, the LHY correction was only recently measured quantitatively using the sophisticated experimental techniques offered by ultracold atomic gases~\cite{navon2010,navon2011}. Moreover,  LHY physics was observed in a number of cold gas experiments~\cite{altmeyer2007, shin2008, papp2008}. A main reason for the difficulty of probing the LHY term is that it typically constitutes a small contribution to a dominant mean-field term.

Recently, it was pointed out that quantum fluctuations can stabilize a Bose-Bose mixture, which would otherwise collapse under attractive mean-field forces~\cite{petrov2015}. The resulting self-bound droplets were observed both without confinement~\cite{semeghini2018}, and with confinement in one~\cite{cabrera2018} or two dimensions~\cite{cheiney2018}. Similar observations were made in dipolar condensates~\cite{kadau2016, ferrier2016observation, schmitt2016, chomaz2016}. While these experiments show the presence of quantum fluctuations, the existence of various competing mean-field and quantum fluctuation energies of the same magnitude, complicates a direct measurement of the LHY term.

In this Letter, we propose a novel approach to studying quantum fluctuations. By  tuning the interactions and atom numbers in a two-component Bose-Einstein condensate (BEC), it is possible to realize a fluid where mean-field interactions are entirely absent. This leaves the LHY correction  as the \emph{only} relevant interaction energy for weak coupling, and we denote such a system a  LHY fluid. We develop an effective single-component framework based on a generalized Gross-Piteavskii equation (GPE), which is shown to accurately describe the LHY fluid by comparing with full numerical two-component simulations. Using the one-component framework, we calculate a number of relevant parameters describing the fluid. Moreover, we show how two powerful measurement techniques available to cold gas experiments, collective oscillations and radio-frequency (RF) spectroscopy, can be used to accurately probe LHY physics. Finally, we show that our results are robust towards considerable deviations in atom numbers and interaction strengths away from the conditions for realizing the ideal LHY fluid.

A two-component BEC at zero temperture is, excluding LHY terms,  described by the mean-field energy functional
\begin{align}
E_\text{MF}= \int \bigg( \sum_{i} \bigg[\frac{\hbar^2|\nabla \Psi_i  |^2}{2m_i}  +  V_i n_i\bigg] +\frac12\sum_{ij}g_{ij}n_in_j \bigg) d\mathbf{r}.
\label{eq:energy_func}
\end{align}
Here, the subscript $i=1,2$ refers to the two components, $\Psi_i (\textbf{r}) = N_i^{1/2} \psi_i (\textbf{r})$ is the condensate wave function, $g_{ij} = 2 \pi \hbar^2 a_{ij}(m_i+m_j)/(m_im_j)$ is the coupling constant between components $i$ and $j$ with $a_{ij}$ corresponding to the scattering length. The density of component $i$ is $n_i(\textbf{r})=|\Psi_i(\textbf{r})|^2$. For simplicity, we choose equal masses $m = m_1 = m_2$ of the two components and a symmetric harmonic trap $V_1(\textbf{r}) = V_2(\textbf{r}) = m\omega_0^2 r^2 /2$. Stability requires $g_{11}>0$ and $g_{22}>0$.

It is easy to show that for $g_{12}=-\sqrt{g_{11}g_{22}}$, one eigenvalue of the quadratic form $\sum_{ij}g_{ij}n_in_j/2$ in Eq.~\eqref{eq:energy_func} is zero whereas the other is positive. The eigenvector associated with the zero eigenvalue corresponds to $n_2=\sqrt{g_{11}/g_{22}}n_1$, which shows that the mean-field energy vanishes for this density ratio, and that any deviation away from this is energetically costly. Thus, by choosing $g_{12}=-\sqrt{g_{11}g_{22}}$ and atom numbers such that $N_2/N_1 = \sqrt{g_{11}/g_{22}}$, the mean-field terms of Eq.~\eqref{eq:energy_func} cancel entirely and the condensate wave functions  take the form $\Psi_2 = \Psi_1 (g_{11}/g_{22})^{1/4}$. The system will   behave as if it is non-interacting at the mean-field level, and any perturbation to one component results in a restoring force towards $\Psi_2 \propto \Psi_1$. This motivates using the ansatz $\Psi_2 = \Psi_1 (g_{11}/g_{22})^{1/4}$.
  
The contribution from quantum fluctuations to the local energy density in a Bose mixture of equal masses $m$ reads~\cite{larsen1963, petrov2015}
\begin{align}
\frac{E_\text{LHY}}{V} & = \frac{32\sqrt{2\pi}}{15} \frac{\hbar^2}{m} \sum_{\pm}( a_{11}n_1 + a_{22}n_2\pm\kappa)^{5/2},
\label{eq:LHY_term_2comp}
\end{align}
where $\kappa=[(a_{11}n_1 - a_{22}n_2)^2 + 4a_{12}^2n_1n_2]^{1/2}$. For $a_{12}=-\sqrt{a_{11}a_{22}}$ and $n_1/n_2 = \sqrt{a_{22}/{a_{11}}}$, this 
 reduces to
\begin{align}
\frac{E_\text{LHY}}{V} = \frac{256\sqrt{\pi}}{15} \frac{\hbar^2}{m} \left( n |a_{12}| \right)^{5/2},
\end{align}
where $n = n_1+n_2$. 

Including this term in  Eq.~\eqref{eq:energy_func} and  defining  $|\Psi|^2 = |\Psi_1|^2+|\Psi_2|^2$ with 
$\Psi_2 = \Psi_1 (g_{11}/g_{22})^{1/4}$ yields the one-component energy functional~\footnote{A gradient expansion for a one-component BEC yields higher order derivative terms~\cite{Braaten1997}. They will however be subleading compared to the kinetic term in Eq.~\eqref{eq:LHYEnergy}.}
\begin{align}
E=
\int \bigg[ \frac{\hbar^2|\nabla \Psi|^2 }{2m}   +  V|\Psi|^2
 + \frac{256\sqrt{\pi}}{15} \frac{\hbar^2}{m} |a_{12}|^{5/2} |\Psi|^5 \bigg] d\textbf{r}.
 \label{eq:LHYEnergy}
\end{align}
Note that there are \emph{no mean-field terms} in this energy functional, so interaction effects are given by the next order LHY fluctuation term for weak interactions. We therefore denote this system as a LHY fluid. The corresponding GPE is
\begin{align}
\mu \Psi  = \left[ - \frac{\hbar^2}{2m} \nabla^2 + V({\bf r}) + \frac{128\sqrt{\pi}}{3} \frac{\hbar^2}{m} |a_{12}|^{5/2} |\Psi|^3 \right] \Psi.
\label{eq:LHYGPE}
\end{align}
This is analogous to the usual one-component GPE, but with the non-linear mean-field $|\Psi|^2$ term replaced by a $|\Psi|^3$ term. It follows from Eq.~\eqref{eq:LHYGPE} that the relevant dimensionless parameter for the interaction strength is $N^{3/2}|a_{12}/a_\text{ho}|^{5/2}$, where $N=N_1+N_2$ is the total number of atoms and $a_\text{ho} = \sqrt{\hbar/m\omega_0}$ is the harmonic oscillator length. This should be compared with the  parameter $Na_{11}/a_\text{ho}$ for a regular BEC.
The simplicity of Eq.~\eqref{eq:LHYGPE} allows for detailed analytical studies of the system, in contrast to multi-component systems which in general are difficult to describe analytically.

Note that the LHY contribution to the chemical potentials of the two components, $\mu_j=\partial E_\text{LHY}/\partial N_{j}$, differs for the two components except for $a_{11} = a_{22}$, as follows from Eq.~\eqref{eq:LHY_term_2comp}. Hence, the LHY term will break  the proportionality $\Psi_2 = \Psi_1 (g_{11}/g_{22})^{1/4}$, which is assumed when deriving Eq.~\eqref{eq:LHYGPE}. However, the mean-field forces will tend to restore this proportionality, and as we shall  see, the one-component framework is  accurate for an describing mixtures with $a_{11} \neq a_{22}$. 
 
\begin{figure}[tb]
	\centering
	\includegraphics{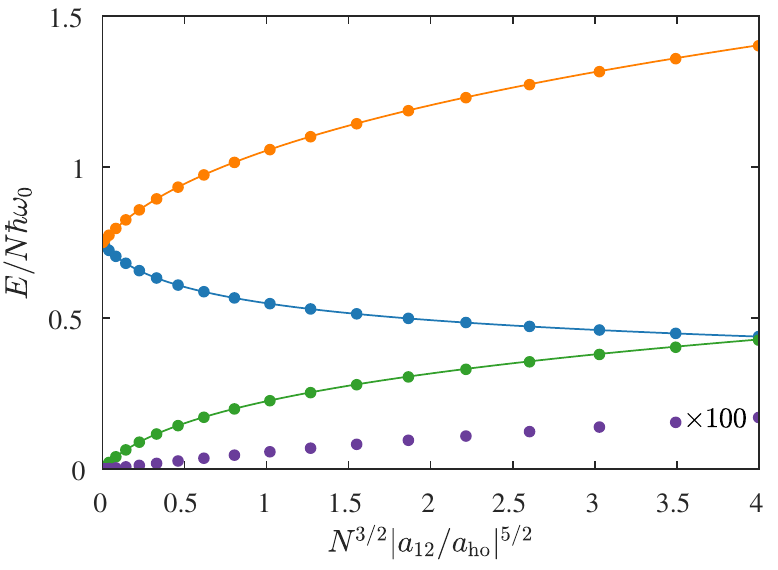}
	\caption{Numerically obtained energies of the LHY fluid.  The lines refer to one-component calculations, and the points to two-component calculations. From top to bottom: orange is the potential energy, blue is the kinetic energy, green is the LHY correction, and purple is the mean-field energy, which only exists in the two-component framework, multiplied by a factor of $100$ for visibility.}
\label{fig:energies}
\end{figure}

To verify the validity of the one-component description,  we compare  the ground-state energy obtained from Eq.~\eqref{eq:LHYGPE} with that 
 using a full two-component GPE formalism with mean-field and LHY terms included~\cite{SM}. For concreteness, we consider the recently studied $^{39}$K spin mixture in states $\ket{1} = \ket{F=1,m_F=-1}$ and $\ket{2} =\ket{1,0}$~\cite{lysebo2010, semeghini2018, cabrera2018, cheiney2018}. At a magnetic field of $\SI{56.8}{G}$, the scattering lengths
  of the system fulfill $|a_{12}| = \sqrt{a_{11}a_{22}} = 53 a_0$ with $a_{22}/a_{11} = 2.5$, where $a_0$ is the Bohr radius~\cite{lysebo2010}. The calculations are performed by varying $\omega_0/2\pi$ from 0 up to $\SI{100}{Hz}$ with a total atom number $N=10^5$ using a numerical toolbox~\cite{antoine2014,antoine2015}. The ground-state energies as a function of interaction strength are shown in Fig.~\ref{fig:energies}. The one- and two-component calculations yield essentially the same results confirming that Eq.~\eqref{eq:LHYGPE} provides an accurate description of the system. The results also show that the mean-field energy in the two-component theory is much smaller than all the other energies, consistent with our approach. The one-component framework indeed remains accurate for interaction strengths beyond those available in typical experiments (see supplementary material~\cite{SM}).

We now use the one-component framework to derive a number of relevant properties of the system. For strong interactions $N^{3/2}(|a_{12}|/a_\text{ho})^{5/2} \gg 1$, the local 
density is well described within the Thomas-Fermi approximation giving
\begin{align}
n(r) = |\Psi(r)|^2 = n_0 (1-r^2/R_\text{LHY}^2)^{2/3},
\label{eq:TFapp}
\end{align}
where $R_\text{LHY}$ is the  fluid radius. Using $V(R_\text{LHY}) = \mu$ yields
\begin{align}
\frac{\mu}{\hbar \omega_0} = \frac{A^2}{2} \left( N^{3/2} \left| \frac{a_{12}}{a_\text{ho}} \right|^{5/2} \right)^{4/13},
\end{align}
 where $A = [91^3 \Gamma(1/3)^9/3^5\pi^8]^{1/13} \approx 1.815$. Similarly, 
 \begin{align}
\frac{R_\text{LHY}}{a_\text{ho}} = A \left( N^{3/2} \left| \frac{a_{12}}{a_\text{ho}} \right|^{5/2} \right)^{2/13}.
\end{align}
Using $\mu = \partial E / \partial N$ and $\mu \propto N^{6/13}$, we obtain $E/N = 13\mu/19$ for the energy per particle.
The central density of the fluid is
\begin{align}
n_0 |a_{12}|^3 =  \frac{A^{4/3}3^{2/3}}{32(2\pi)^{1/3}} \left( N \left| \frac{a_{12}}{a_\text{ho}} \right|^6 \right)^{4/13}.
\label{eq:Density}
\end{align}
This can also be used to confirm the accuracy of our weak coupling  theory, valid for $n_0 |a_{12}|^3 \ll 1$. For typical values $N=10^5$ and $a_{12} = 53 a_0$, an interaction strength of $N^{3/2}|a_{12}/a_\text{ho}|^{5/2} \approx 1.7 \times 10^4$ would be required to reach 1\% of this limiting criterion ($n_0 |a_{12}|^3 = 0.01$). This illustrates that all our results are  in the weakly interacting limit $n_0 |a_{12}|^3\ll 1$.

We also introduce a healing length $\xi_\text{LHY}$  providing the typical length scale of density variations of the LHY fluid. 
 It is determined by a competition between the kinetic energy  and the interaction energy. Equating $\hbar^2/2m\xi_\text{LHY}^2=(128\sqrt{\pi}\hbar^2/3m)n^{3/2}|a_{12}|^{5/2}$
 yields
\begin{align}
\xi_\text{LHY}^2 = \frac{3}{256 \sqrt{\pi} |a_{12}|^{5/2} n^{3/2}}.
\end{align}

One of the most powerful techniques for investigation of cold atomic gases is collective excitations~\cite{jin1996, stamper1998, dalfovo1999, pethick2002, altmeyer2007, Riedl2008}. We now show how the simplest collective excitation, the monopole breathing mode,  can be used to probe the LHY correlations. To do so, we introduce the generic wave function~\cite{pethick2002}
\begin{align}
\Psi (r) = \frac{BN^{1/2}}{R^{3/2}} f(r/R) e^{i \phi},
\label{eq:Monopoleansatz}
\end{align}
where $f$ is an arbitrary real function, $R$ is a radius, $\phi$ is a phase and $B$ is a normalization constant. Inserting Eq.~\eqref{eq:Monopoleansatz} into 
Eq.~\eqref{eq:LHYEnergy} gives the energy
\begin{align}
E =  E_\text{flow} + E_\text{pot} + E_\text{zp} + E_\text{LHY} = E_\text{flow} + U(R).
\end{align}
Here,  $E_\text{flow} = \hbar^2/2m \int d\textbf{r} |\Psi|^2 (\nabla\phi)^2$ is the kinetic energy of the particle currents, 
$ E_\text{pot} = m\omega_0^2/2 \int d\textbf{r} r^2 |\Psi|^2 \propto R^{2}$ is the potential energy, 
$E_\text{LHY} = 256 \sqrt{\pi}\hbar^2 |a_{12}|^2 /15m \int d\textbf{r} |\Psi|^5 \propto R^{-9/2}$ is the LHY correction, and 
$E_\text{zp} = \hbar^2/2m \int d\textbf{r} (d|\Psi|/dr)^2 \propto R^{-2}$ the zero-point kinetic energy.

Consider first the equilibrium  case $R=R_0$, where $E_\text{flow} = 0$ and $dU/dR |_{R=R_0} = 0$. The energy terms are proportional to powers of $R$ from which follows a virial theorem
\begin{align}
R \frac{dU}{dR} \bigg |_{R=R_0} =  2 E_\text{pot} -2 E_\text{zp} - \frac{9}{2} E_\text{LHY} = 0.
\label{eq:virial}
\end{align}
Now, we move on to dynamics by considering a time-dependent $R$. The corresponding particle velocity is  $v(r) = r \dot R/R$, and  $E_\text{flow} = m_\text{e} \dot R^2/2$ where $m_\text{e} = Nm \langle r^2 \rangle/R^2$ is an effective mass, and $\langle r^2 \rangle$ is the mean-square radius of the fluid~\cite{pethick2002}. For a harmonic oscillator, the effective mass is $m_\text{e} = 2 E_\text{pot} / \omega_0^2R^2$. Conservation of the total energy $m_\text{e} \dot R^2/2 + U(R)$ gives the  equation of motion $m_\text{e} \ddot R = - \partial U(R)/\partial R $.

Expanding the effective potential  $U(R)$ to second order around equilibrium gives  $U(R) = U(R_0) + C(R - R_0)^2/2$, where $C = d^2 U(R) / dR^2$ is a constant describing the restoring force towards equilibrium. The monopole oscillation frequency is therefore $\omega^2 = C/m_\text{e}$. By calculating $R^2 d^2 U(R) / dR^2$ and using Eq.~\eqref{eq:virial}, we obtain
\begin{align}
\omega^2 = \omega_0^2 \left( 4 + \frac{45}{8} \frac{E_\text{LHY}}{E_\text{pot}} \right).
\label{eq:monopole}
\end{align}
This expression allows for a straightforward evaluation of the monopole frequency at various interaction strengths.

In the Thomas-Fermi limit we can set $E_\text{zp}=0$ in Eqs.~(\ref{eq:virial}--\ref{eq:monopole}) obtaining  $\omega/ \omega_0 = \sqrt{13/2}$.
The frequency change  compared to the non-interacting value $\omega/ \omega_0 = 2$ is thus  more than twice as large than for a regular BEC  where $\omega/ \omega_0 = \sqrt{5}$ in the Thomas-Fermi limit~\cite{dalfovo1999}.
For weak interaction, $E_\text{LHY}$ and $E_\text{pot}$ can be calculated using the harmonic oscillator ground-state wave function, and Eq.~\eqref{eq:monopole} yields 
\begin{align}
\frac\omega\omega_0= 2 + \pi^{-7/4} \frac{64 \sqrt{2}}{5 \sqrt{5}} N^{3/2} \left| \frac{a_{12}}{a_\text{ho}} \right|^{5/2} .
\label{eq:monopole_weak}
\end{align}

To evaluate the monopole frequency for intermediate interactions, we use the energies obtained from the one-component ground-state calculations shown in Fig.~\ref{fig:energies} combined with Eq.~\eqref{eq:monopole}. As a check, we have also performed dynamical simulations using the full two-component description~\cite{SM}.

\begin{figure}[tb]
	\centering
	\includegraphics{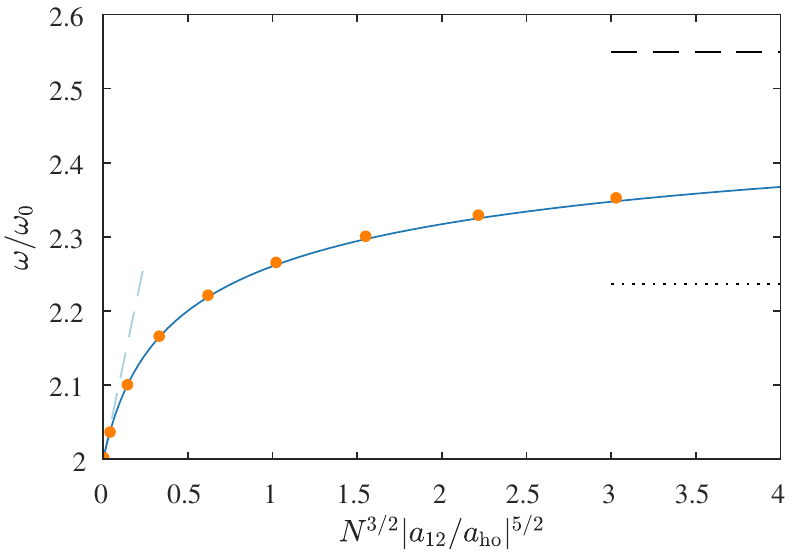}
	\caption{Monopole  frequency of the LHY fluid. The light blue dashed line is Eq.~\eqref{eq:monopole_weak}, valid for weak interactions. The black 
	dashed and dotted lines give the Thomas-Fermi results $\omega/ \omega_0 = \sqrt{13/2}$ and $\omega/ \omega_0 = \sqrt{5}$ for a
	 LHY fluid and a regular BEC, respectively. The solid blue line is obtained from Eq.~\eqref{eq:monopole} combined with ground-state calculations performed in the one-component framework, while the points are obtained from dynamical two-component simulations.}
\label{fig:osc_freqs}
\end{figure}

In Fig.~\ref{fig:osc_freqs}, we show the monopole frequency as a function of the interaction strength. The one-component framework for the LHY fluid yields essentially the same frequency as the full two-component calculations, demonstrating its accuracy. Importantly, it confirms that the frequency shift of the monopole mode is driven almost exclusively  by the LHY correlations, whereas mean-field forces play a negligible role. Figure~\ref{fig:osc_freqs} also displays how the monopole frequency interpolates between the result of Eq.~\eqref{eq:monopole_weak} for $N^{3/2}|a_{12}/a_\text{ho}|^{5/2}\ll 1$ and $\omega/ \omega_0 = \sqrt{13/2}$ for $N^{3/2}|a_{12}/a_\text{ho}|^{5/2}\gg 1$, which is also shown in the supplemental material for stronger interactions~\cite{SM}. Given the high accuracy of collective mode experiments, we conclude that the LHY correlations can be probed quantitatively by measuring the monopole frequency of a LHY fluid. 

We now show that RF spectroscopy, which has been used extensively to study interaction effects in BECs~\cite{harber2002, papp2008, wild2012, jorgensen2016},
can be used to directly measure the LHY energy by  transferring atoms between components $1$ and $2$. Since a small number of atoms transferred from one component to the other will experience the same mean-field energy in the two states, any shift of the transition frequency is entirely due to the LHY energy term, which differ for the two components. 
\begin{figure}[tb]
	\centering
	\includegraphics{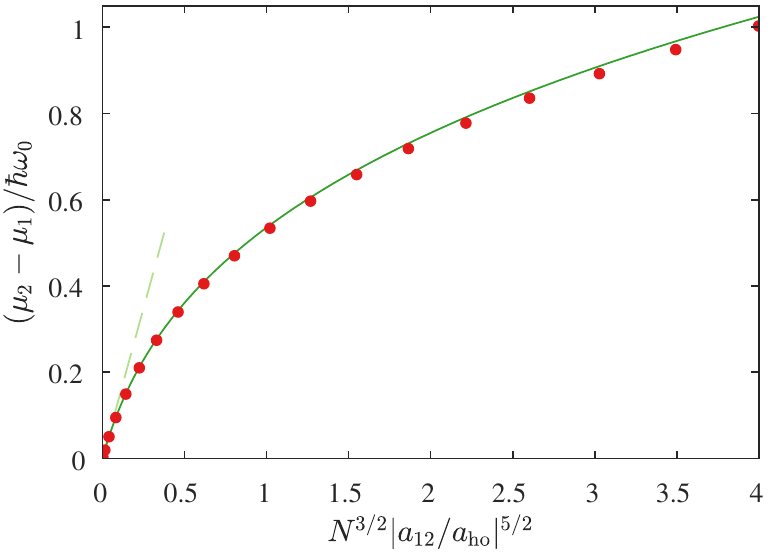}
	\caption{Difference in chemical potential between two components of the LHY fluid, which is observable through RF spectroscopy. The light green dashed line is Eq.~\eqref{eq:mu_shift_weak}, which is valid for weakly coupling. The full green line is obtained from ground-state calculations using the one-component framework combined with Eq.~\eqref{eq:mu_shift}, whereas the points are from the two-component framework.}
\label{fig:chemPotDiff}
\end{figure}
The shift is determined by the difference $\mu_1-\mu_2$ in chemical potentials of the two states~\cite{SM}. Assuming that $|\Psi|^2/N = |\Psi_1|^2/N_1 = |\Psi_2|^2/N_2$
yields the shift~\cite{SM}
\begin{align}
\frac{\mu_2 - \mu_1}{ \mu_\text{LHY} } = \frac{a_{22} - a_{11}}{|a_{12}|},
\label{eq:mu_shift}
\end{align}
where $ \mu_\text{LHY}  = (128\sqrt{\pi}\hbar^2|a_{12}|^{5/2}/3m)\expval{|\Psi(\mathbf r)|^3}{\Psi}/\bra{\Psi}\ket{\Psi}$
is the expectation value of the LHY interaction energy in the one-component framework. It is thus straightforward to obtain the shift by calculating $\mu_\text{LHY} $.

To obtain the shift in the weakly interacting limit, the value $\mu_\text{LHY} $ is calculated assuming the harmonic oscillator ground-state wave function
\begin{align}
\frac{\mu_2 - \mu_1}{\hbar \omega_0} = \pi^{-7/4} \frac{512\sqrt{5}}{75\sqrt{2}} \frac{a_{22} - a_{11}}{|a_{12}|} N^{3/2} \left| \frac{a_{12}}{a_\text{ho}} \right|^{5/2}.
\label{eq:mu_shift_weak}
\end{align}

To evaluate this, we again consider $^{39}$K in the spin states given above. The resulting shift using Eq.~\eqref{eq:mu_shift_weak} is shown in Fig.~\ref{fig:chemPotDiff}. To calculate the shift for stronger interactions, the ground-state calculations for both the one- and the two-component frameworks shown in Fig.~\ref{fig:energies} are used. For the one-component framework, $\mu_\text{LHY}$ is extracted and Eq.~\eqref{eq:mu_shift} is used to calculate the frequency shift. For the two-component framework, the difference of the numerically obtained chemical potentials is shown. 

Figure~\ref{fig:chemPotDiff} shows that the quantum fluctuations clearly result in a difference of chemical potentials, and the resonance frequency for transferring atoms is shifted accordingly. Since the two components do not have identical atom numbers $N_1 \neq N_2$, a resonant RF pulse will not transfer the same number of atoms between the two components, and it will therefore result in a deviation from the ratio $N_2/N_1 = \sqrt{a_{11}/a_{22}}$. RF spectroscopy is thereby capable of directly measuring the energy contribution from quantum fluctuations in a LHY fluid.

\begin{figure}[tb]
	\centering
	\includegraphics{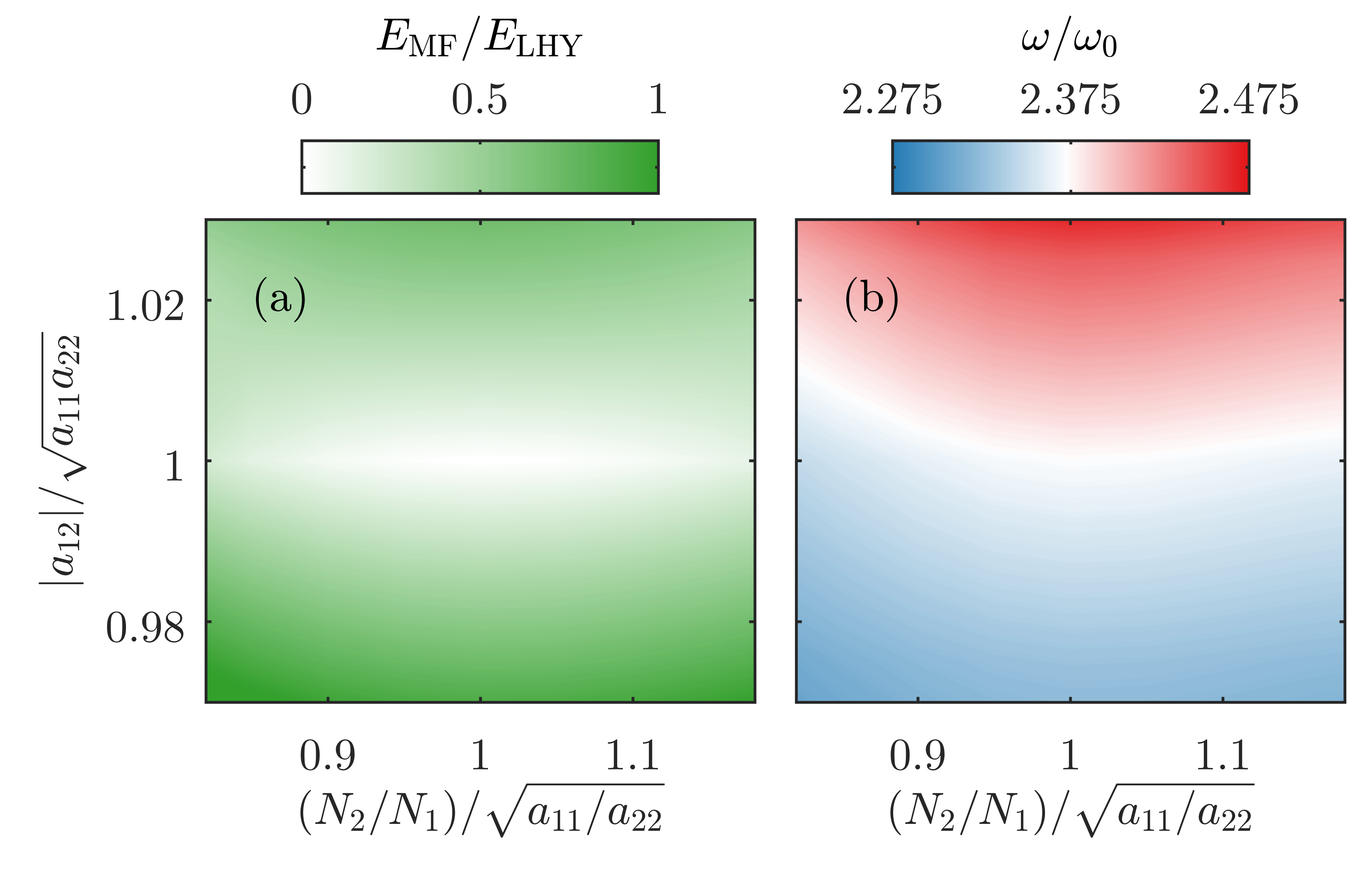}
	\caption{Properties of the LHY fluid when deviating from the criteria $N_2/N_1 = \sqrt{a_{11}/a_{22}}$ and $a_{12} = - \sqrt{a_{11}a_{22}}$ for $N^{3/2}(\sqrt{a_{11}a_{22}}/a_\text{ho})^{5/2} = 4$. (a) The mean-field energy relative to the LHY energy. (b) The monopole oscillation frequency $\omega/\omega_0$.}
\label{fig:errorFig}
\end{figure}

 It is experimentally challenging to perfectly match the criteria $N_2/N_1 = \sqrt{a_{11}/a_{22}}$ and $a_{12} = - \sqrt{a_{11}a_{22}}$ as required. An important question
 for the feasibility of realizing a LHY fluid is therefore the effect of small deviations from these values. To investigate this,  we perform full two-component ground-state calculations with constant $a_{11}$, $a_{22}$, $a_\text{ho}$, and $N=10^5$, corresponding to $N^{3/2}(\sqrt{a_{11}a_{22}}/a_\text{ho})^{5/2} = 4$,  but with various values of $a_{12}$ and $N_1/N_2$. Figure~\ref{fig:errorFig}(a) shows that the mean-field energy is small compared to the LHY energy so quantum fluctuations are the dominating source of interaction effects, even for relatively large deviations. In Fig.~\ref{fig:errorFig}(b), we  show the monopole oscillation frequency, obtained from dynamical two-component simulations. It remains significantly larger than the frequency $\omega/\omega_0 =\sqrt{5} \approx 2.236$ of a regular BEC in the Thomas-Fermi limit for a broad range of parameters. In both cases, Fig.~\ref{fig:errorFig} shows that the LHY fluid is more susceptible to deviations in scattering length than to deviations in relative atom numbers. This can be understood from the fact that if an atom of type 1 is added to a mixture fulfilling $N_2/N_1 = \sqrt{a_{11}/a_{22}}$ and $a_{12} = -\sqrt{a_{11} a_{22}}$, the net contribution to mean-field energy is $\sim n_1 g_{11} + n_2 g_{12} = 0$.  However, changing the scattering length away from $a_{12} = -\sqrt{a_{11} a_{22}}$ leads to an imbalance in net mean-field energy, which quickly becomes comparable to the LHY energy. 
 
These results can be used to discuss the experimental feasibility of realizing a LHY fluid. As argued above, it is mainly important to fulfill the scattering length criterion $a_{12} = -\sqrt{a_{11} a_{22}}$. For the $^{39}$K system discussed in this work, the range $0.97$--$1.03$ of $a_{12}/\sqrt{a_{11} a_{22}}$ corresponds to a magnetic field window of approximately $\SI{250}{mG}$. Within this range, which is well within the precision of ultracold gas experiments, it is thus possible to realize a fluid where quantum 
fluctuations are stronger than mean-field interactions. 

In conclusion, we propose the realization of a dilute quantum fluid where quantum fluctuations provide the only relevant interaction. Even though the fluid consists of two distinct components, it is possible to describe it using a one-component framework. We furthermore propose two methods to study the quantum fluctuations: collective oscillations and RF spectroscopy. Finally, we have shown that it is experimentally feasible to realize the fluid, despite typical uncertainties in scattering length and atom numbers. 
The  one-component framework will stimulate further theoretical studies -- in future work, it will be interesting to investigate, e.\,g., the dispersion relation and the expansion properties of the fluid.
 
The realization of the LHY fluid opens up for an unprecedented characterization of quantum fluctuations. Since it is possible to directly study the LHY correction without the presence of predominant mean-field contributions, this system is a promising candidate for observing \emph{beyond} LHY corrections. The next order (in $na^3$) correction to the energy in a Bose gas, $E_\text{WHPS}$, was calculated for a single-component BEC~\cite{Fetter1971}, and recently for a single impurity in a BEC~\cite{christensen2015}. Since it has not been calculated for a two-component BEC so far, we use the single-component result as a guide, which gives $E_\text{WHPS}/E_\text{LHY} \approx 4.1 \sqrt{na^3} \ln (na^3)$~\cite{Fetter1971}. Using Eq.~(\ref{eq:Density}) with $N=10^5$, $a_{12} = 53 a_0$, and $\omega_0/2\pi = \SI{1}{kHz}$, corresponding to $N^{3/2}|a_{12}/a_\text{ho}|^{5/2} \approx 72$, yields $E_\text{WHPS}/E_\text{LHY} \approx -0.5$, showing that beyond-LHY physics indeed is experimentally accessible. It would be interesting yet challenging to calculate this term for a two-component mixture, building, for instance, on the diagrammatic results of Refs.~\cite{christensen2015,Utesov2018}.

The LHY fluid constitutes a compelling system, located between two interaction limits of Bose gases: mean-field interactions and highly correlated Bose gases. It thereby provides a stepping stone towards a better understanding of new exotic systems. 

\bibliography{JorgensenBib}

\begin{thebibliography}{32}%
\makeatletter
\providecommand \@ifxundefined [1]{%
 \@ifx{#1\undefined}
}%
\providecommand \@ifnum [1]{%
 \ifnum #1\expandafter \@firstoftwo
 \else \expandafter \@secondoftwo
 \fi
}%
\providecommand \@ifx [1]{%
 \ifx #1\expandafter \@firstoftwo
 \else \expandafter \@secondoftwo
 \fi
}%
\providecommand \natexlab [1]{#1}%
\providecommand \enquote  [1]{``#1''}%
\providecommand \bibnamefont  [1]{#1}%
\providecommand \bibfnamefont [1]{#1}%
\providecommand \citenamefont [1]{#1}%
\providecommand \href@noop [0]{\@secondoftwo}%
\providecommand \href [0]{\begingroup \@sanitize@url \@href}%
\providecommand \@href[1]{\@@startlink{#1}\@@href}%
\providecommand \@@href[1]{\endgroup#1\@@endlink}%
\providecommand \@sanitize@url [0]{\catcode `\\12\catcode `\$12\catcode
  `\&12\catcode `\#12\catcode `\^12\catcode `\_12\catcode `\%12\relax}%
\providecommand \@@startlink[1]{}%
\providecommand \@@endlink[0]{}%
\providecommand \url  [0]{\begingroup\@sanitize@url \@url }%
\providecommand \@url [1]{\endgroup\@href {#1}{\urlprefix }}%
\providecommand \urlprefix  [0]{URL }%
\providecommand \Eprint [0]{\href }%
\providecommand \doibase [0]{http://dx.doi.org/}%
\providecommand \selectlanguage [0]{\@gobble}%
\providecommand \bibinfo  [0]{\@secondoftwo}%
\providecommand \bibfield  [0]{\@secondoftwo}%
\providecommand \translation [1]{[#1]}%
\providecommand \BibitemOpen [0]{}%
\providecommand \bibitemStop [0]{}%
\providecommand \bibitemNoStop [0]{.\EOS\space}%
\providecommand \EOS [0]{\spacefactor3000\relax}%
\providecommand \BibitemShut  [1]{\csname bibitem#1\endcsname}%
\let\auto@bib@innerbib\@empty
\bibitem [{\citenamefont {Lee}\ \emph {et~al.}(1957)\citenamefont {Lee},
  \citenamefont {Huang},\ and\ \citenamefont {Yang}}]{lee1957}%
  \BibitemOpen
  \bibfield  {author} {\bibinfo {author} {\bibfnamefont {T.~D.}\ \bibnamefont
  {Lee}}, \bibinfo {author} {\bibfnamefont {K.}~\bibnamefont {Huang}}, \ and\
  \bibinfo {author} {\bibfnamefont {C.~N.}\ \bibnamefont {Yang}},\ }\href@noop
  {} {\bibfield  {journal} {\bibinfo  {journal} {Phys. Rev.}\ }\textbf
  {\bibinfo {volume} {106}},\ \bibinfo {pages} {1135} (\bibinfo {year}
  {1957})}\BibitemShut {NoStop}%
\bibitem [{\citenamefont {Navon}\ \emph {et~al.}(2010)\citenamefont {Navon},
  \citenamefont {Nascimb{\`e}ne}, \citenamefont {Chevy},\ and\ \citenamefont
  {Salomon}}]{navon2010}%
  \BibitemOpen
  \bibfield  {author} {\bibinfo {author} {\bibfnamefont {N.}~\bibnamefont
  {Navon}}, \bibinfo {author} {\bibfnamefont {S.}~\bibnamefont
  {Nascimb{\`e}ne}}, \bibinfo {author} {\bibfnamefont {F.}~\bibnamefont
  {Chevy}}, \ and\ \bibinfo {author} {\bibfnamefont {C.}~\bibnamefont
  {Salomon}},\ }\href@noop {} {\bibfield  {journal} {\bibinfo  {journal}
  {Science}\ }\textbf {\bibinfo {volume} {328}},\ \bibinfo {pages} {729}
  (\bibinfo {year} {2010})}\BibitemShut {NoStop}%
\bibitem [{\citenamefont {Navon}\ \emph {et~al.}(2011)\citenamefont {Navon},
  \citenamefont {Piatecki}, \citenamefont {G{\"u}nter}, \citenamefont {Rem},
  \citenamefont {Nguyen}, \citenamefont {Chevy}, \citenamefont {Krauth},\ and\
  \citenamefont {Salomon}}]{navon2011}%
  \BibitemOpen
  \bibfield  {author} {\bibinfo {author} {\bibfnamefont {N.}~\bibnamefont
  {Navon}}, \bibinfo {author} {\bibfnamefont {S.}~\bibnamefont {Piatecki}},
  \bibinfo {author} {\bibfnamefont {K.}~\bibnamefont {G{\"u}nter}}, \bibinfo
  {author} {\bibfnamefont {B.}~\bibnamefont {Rem}}, \bibinfo {author}
  {\bibfnamefont {T.~C.}\ \bibnamefont {Nguyen}}, \bibinfo {author}
  {\bibfnamefont {F.}~\bibnamefont {Chevy}}, \bibinfo {author} {\bibfnamefont
  {W.}~\bibnamefont {Krauth}}, \ and\ \bibinfo {author} {\bibfnamefont
  {C.}~\bibnamefont {Salomon}},\ }\href@noop {} {\bibfield  {journal} {\bibinfo
   {journal} {Phys. Rev. Lett.}\ }\textbf {\bibinfo {volume} {107}},\ \bibinfo
  {pages} {135301} (\bibinfo {year} {2011})}\BibitemShut {NoStop}%
\bibitem [{\citenamefont {Altmeyer}\ \emph {et~al.}(2007)\citenamefont
  {Altmeyer}, \citenamefont {Riedl}, \citenamefont {Kohstall}, \citenamefont
  {Wright}, \citenamefont {Geursen}, \citenamefont {Bartenstein}, \citenamefont
  {Chin}, \citenamefont {Denschlag},\ and\ \citenamefont
  {Grimm}}]{altmeyer2007}%
  \BibitemOpen
  \bibfield  {author} {\bibinfo {author} {\bibfnamefont {A.}~\bibnamefont
  {Altmeyer}}, \bibinfo {author} {\bibfnamefont {S.}~\bibnamefont {Riedl}},
  \bibinfo {author} {\bibfnamefont {C.}~\bibnamefont {Kohstall}}, \bibinfo
  {author} {\bibfnamefont {M.}~\bibnamefont {Wright}}, \bibinfo {author}
  {\bibfnamefont {R.}~\bibnamefont {Geursen}}, \bibinfo {author} {\bibfnamefont
  {M.}~\bibnamefont {Bartenstein}}, \bibinfo {author} {\bibfnamefont
  {C.}~\bibnamefont {Chin}}, \bibinfo {author} {\bibfnamefont {J.~H.}\
  \bibnamefont {Denschlag}}, \ and\ \bibinfo {author} {\bibfnamefont
  {R.}~\bibnamefont {Grimm}},\ }\href@noop {} {\bibfield  {journal} {\bibinfo
  {journal} {Phys. Rev. Lett.}\ }\textbf {\bibinfo {volume} {98}},\ \bibinfo
  {pages} {040401} (\bibinfo {year} {2007})}\BibitemShut {NoStop}%
\bibitem [{\citenamefont {Shin}\ \emph {et~al.}(2008)\citenamefont {Shin},
  \citenamefont {Schirotzek}, \citenamefont {Schunck},\ and\ \citenamefont
  {Ketterle}}]{shin2008}%
  \BibitemOpen
  \bibfield  {author} {\bibinfo {author} {\bibfnamefont {Y.-i.}\ \bibnamefont
  {Shin}}, \bibinfo {author} {\bibfnamefont {A.}~\bibnamefont {Schirotzek}},
  \bibinfo {author} {\bibfnamefont {C.~H.}\ \bibnamefont {Schunck}}, \ and\
  \bibinfo {author} {\bibfnamefont {W.}~\bibnamefont {Ketterle}},\ }\href@noop
  {} {\bibfield  {journal} {\bibinfo  {journal} {Phys. Rev. Lett.}\ }\textbf
  {\bibinfo {volume} {101}},\ \bibinfo {pages} {070404} (\bibinfo {year}
  {2008})}\BibitemShut {NoStop}%
\bibitem [{\citenamefont {Papp}\ \emph {et~al.}(2008)\citenamefont {Papp},
  \citenamefont {Pino}, \citenamefont {Wild}, \citenamefont {Ronen},
  \citenamefont {Wieman}, \citenamefont {Jin},\ and\ \citenamefont
  {Cornell}}]{papp2008}%
  \BibitemOpen
  \bibfield  {author} {\bibinfo {author} {\bibfnamefont {S.}~\bibnamefont
  {Papp}}, \bibinfo {author} {\bibfnamefont {J.}~\bibnamefont {Pino}}, \bibinfo
  {author} {\bibfnamefont {R.}~\bibnamefont {Wild}}, \bibinfo {author}
  {\bibfnamefont {S.}~\bibnamefont {Ronen}}, \bibinfo {author} {\bibfnamefont
  {C.~E.}\ \bibnamefont {Wieman}}, \bibinfo {author} {\bibfnamefont {D.~S.}\
  \bibnamefont {Jin}}, \ and\ \bibinfo {author} {\bibfnamefont {E.~A.}\
  \bibnamefont {Cornell}},\ }\href@noop {} {\bibfield  {journal} {\bibinfo
  {journal} {Phys. Rev. Lett.}\ }\textbf {\bibinfo {volume} {101}},\ \bibinfo
  {pages} {135301} (\bibinfo {year} {2008})}\BibitemShut {NoStop}%
\bibitem [{\citenamefont {Petrov}(2015)}]{petrov2015}%
  \BibitemOpen
  \bibfield  {author} {\bibinfo {author} {\bibfnamefont {D.}~\bibnamefont
  {Petrov}},\ }\href@noop {} {\bibfield  {journal} {\bibinfo  {journal} {Phys.
  Rev. Lett.}\ }\textbf {\bibinfo {volume} {115}},\ \bibinfo {pages} {155302}
  (\bibinfo {year} {2015})}\BibitemShut {NoStop}%
\bibitem [{\citenamefont {Semeghini}\ \emph {et~al.}(2018)\citenamefont
  {Semeghini}, \citenamefont {Ferioli}, \citenamefont {Masi}, \citenamefont
  {Mazzinghi}, \citenamefont {Wolswijk}, \citenamefont {Minardi}, \citenamefont
  {Modugno}, \citenamefont {Modugno}, \citenamefont {Inguscio},\ and\
  \citenamefont {Fattori}}]{semeghini2018}%
  \BibitemOpen
  \bibfield  {author} {\bibinfo {author} {\bibfnamefont {G.}~\bibnamefont
  {Semeghini}}, \bibinfo {author} {\bibfnamefont {G.}~\bibnamefont {Ferioli}},
  \bibinfo {author} {\bibfnamefont {L.}~\bibnamefont {Masi}}, \bibinfo {author}
  {\bibfnamefont {C.}~\bibnamefont {Mazzinghi}}, \bibinfo {author}
  {\bibfnamefont {L.}~\bibnamefont {Wolswijk}}, \bibinfo {author}
  {\bibfnamefont {F.}~\bibnamefont {Minardi}}, \bibinfo {author} {\bibfnamefont
  {M.}~\bibnamefont {Modugno}}, \bibinfo {author} {\bibfnamefont
  {G.}~\bibnamefont {Modugno}}, \bibinfo {author} {\bibfnamefont
  {M.}~\bibnamefont {Inguscio}}, \ and\ \bibinfo {author} {\bibfnamefont
  {M.}~\bibnamefont {Fattori}},\ }\href@noop {} {\bibfield  {journal} {\bibinfo
   {journal} {Phys. Rev. Lett.}\ }\textbf {\bibinfo {volume} {120}},\ \bibinfo
  {pages} {235301} (\bibinfo {year} {2018})}\BibitemShut {NoStop}%
\bibitem [{\citenamefont {Cabrera}\ \emph {et~al.}(2018)\citenamefont
  {Cabrera}, \citenamefont {Tanzi}, \citenamefont {Sanz}, \citenamefont
  {Naylor}, \citenamefont {Thomas}, \citenamefont {Cheiney},\ and\
  \citenamefont {Tarruell}}]{cabrera2018}%
  \BibitemOpen
  \bibfield  {author} {\bibinfo {author} {\bibfnamefont {C.}~\bibnamefont
  {Cabrera}}, \bibinfo {author} {\bibfnamefont {L.}~\bibnamefont {Tanzi}},
  \bibinfo {author} {\bibfnamefont {J.}~\bibnamefont {Sanz}}, \bibinfo {author}
  {\bibfnamefont {B.}~\bibnamefont {Naylor}}, \bibinfo {author} {\bibfnamefont
  {P.}~\bibnamefont {Thomas}}, \bibinfo {author} {\bibfnamefont
  {P.}~\bibnamefont {Cheiney}}, \ and\ \bibinfo {author} {\bibfnamefont
  {L.}~\bibnamefont {Tarruell}},\ }\href@noop {} {\bibfield  {journal}
  {\bibinfo  {journal} {Science}\ }\textbf {\bibinfo {volume} {359}},\ \bibinfo
  {pages} {301} (\bibinfo {year} {2018})}\BibitemShut {NoStop}%
\bibitem [{\citenamefont {Cheiney}\ \emph {et~al.}(2018)\citenamefont
  {Cheiney}, \citenamefont {Cabrera}, \citenamefont {Sanz}, \citenamefont
  {Naylor}, \citenamefont {Tanzi},\ and\ \citenamefont
  {Tarruell}}]{cheiney2018}%
  \BibitemOpen
  \bibfield  {author} {\bibinfo {author} {\bibfnamefont {P.}~\bibnamefont
  {Cheiney}}, \bibinfo {author} {\bibfnamefont {C.~R.}\ \bibnamefont
  {Cabrera}}, \bibinfo {author} {\bibfnamefont {J.}~\bibnamefont {Sanz}},
  \bibinfo {author} {\bibfnamefont {B.}~\bibnamefont {Naylor}}, \bibinfo
  {author} {\bibfnamefont {L.}~\bibnamefont {Tanzi}}, \ and\ \bibinfo {author}
  {\bibfnamefont {L.}~\bibnamefont {Tarruell}},\ }\href@noop {} {\bibfield
  {journal} {\bibinfo  {journal} {Phys. Rev. Lett.}\ }\textbf {\bibinfo
  {volume} {120}},\ \bibinfo {pages} {135301} (\bibinfo {year}
  {2018})}\BibitemShut {NoStop}%
\bibitem [{\citenamefont {Kadau}\ \emph {et~al.}(2016)\citenamefont {Kadau},
  \citenamefont {Schmitt}, \citenamefont {Wenzel}, \citenamefont {Wink},
  \citenamefont {Maier}, \citenamefont {Ferrier-Barbut},\ and\ \citenamefont
  {Pfau}}]{kadau2016}%
  \BibitemOpen
  \bibfield  {author} {\bibinfo {author} {\bibfnamefont {H.}~\bibnamefont
  {Kadau}}, \bibinfo {author} {\bibfnamefont {M.}~\bibnamefont {Schmitt}},
  \bibinfo {author} {\bibfnamefont {M.}~\bibnamefont {Wenzel}}, \bibinfo
  {author} {\bibfnamefont {C.}~\bibnamefont {Wink}}, \bibinfo {author}
  {\bibfnamefont {T.}~\bibnamefont {Maier}}, \bibinfo {author} {\bibfnamefont
  {I.}~\bibnamefont {Ferrier-Barbut}}, \ and\ \bibinfo {author} {\bibfnamefont
  {T.}~\bibnamefont {Pfau}},\ }\href@noop {} {\bibfield  {journal} {\bibinfo
  {journal} {Nature}\ }\textbf {\bibinfo {volume} {530}},\ \bibinfo {pages}
  {194} (\bibinfo {year} {2016})}\BibitemShut {NoStop}%
\bibitem [{\citenamefont {Ferrier-Barbut}\ \emph {et~al.}(2016)\citenamefont
  {Ferrier-Barbut}, \citenamefont {Kadau}, \citenamefont {Schmitt},
  \citenamefont {Wenzel},\ and\ \citenamefont {Pfau}}]{ferrier2016observation}%
  \BibitemOpen
  \bibfield  {author} {\bibinfo {author} {\bibfnamefont {I.}~\bibnamefont
  {Ferrier-Barbut}}, \bibinfo {author} {\bibfnamefont {H.}~\bibnamefont
  {Kadau}}, \bibinfo {author} {\bibfnamefont {M.}~\bibnamefont {Schmitt}},
  \bibinfo {author} {\bibfnamefont {M.}~\bibnamefont {Wenzel}}, \ and\ \bibinfo
  {author} {\bibfnamefont {T.}~\bibnamefont {Pfau}},\ }\href@noop {} {\bibfield
   {journal} {\bibinfo  {journal} {Phys. Rev. Lett.}\ }\textbf {\bibinfo
  {volume} {116}},\ \bibinfo {pages} {215301} (\bibinfo {year}
  {2016})}\BibitemShut {NoStop}%
\bibitem [{\citenamefont {Schmitt}\ \emph {et~al.}(2016)\citenamefont
  {Schmitt}, \citenamefont {Wenzel}, \citenamefont {B{\"o}ttcher},
  \citenamefont {Ferrier-Barbut},\ and\ \citenamefont {Pfau}}]{schmitt2016}%
  \BibitemOpen
  \bibfield  {author} {\bibinfo {author} {\bibfnamefont {M.}~\bibnamefont
  {Schmitt}}, \bibinfo {author} {\bibfnamefont {M.}~\bibnamefont {Wenzel}},
  \bibinfo {author} {\bibfnamefont {F.}~\bibnamefont {B{\"o}ttcher}}, \bibinfo
  {author} {\bibfnamefont {I.}~\bibnamefont {Ferrier-Barbut}}, \ and\ \bibinfo
  {author} {\bibfnamefont {T.}~\bibnamefont {Pfau}},\ }\href@noop {} {\bibfield
   {journal} {\bibinfo  {journal} {Nature}\ }\textbf {\bibinfo {volume}
  {539}},\ \bibinfo {pages} {259} (\bibinfo {year} {2016})}\BibitemShut
  {NoStop}%
\bibitem [{\citenamefont {Chomaz}\ \emph {et~al.}(2016)\citenamefont {Chomaz},
  \citenamefont {Baier}, \citenamefont {Petter}, \citenamefont {Mark},
  \citenamefont {W{\"a}chtler}, \citenamefont {Santos},\ and\ \citenamefont
  {Ferlaino}}]{chomaz2016}%
  \BibitemOpen
  \bibfield  {author} {\bibinfo {author} {\bibfnamefont {L.}~\bibnamefont
  {Chomaz}}, \bibinfo {author} {\bibfnamefont {S.}~\bibnamefont {Baier}},
  \bibinfo {author} {\bibfnamefont {D.}~\bibnamefont {Petter}}, \bibinfo
  {author} {\bibfnamefont {M.}~\bibnamefont {Mark}}, \bibinfo {author}
  {\bibfnamefont {F.}~\bibnamefont {W{\"a}chtler}}, \bibinfo {author}
  {\bibfnamefont {L.}~\bibnamefont {Santos}}, \ and\ \bibinfo {author}
  {\bibfnamefont {F.}~\bibnamefont {Ferlaino}},\ }\href@noop {} {\bibfield
  {journal} {\bibinfo  {journal} {Phys. Rev. X}\ }\textbf {\bibinfo {volume}
  {6}},\ \bibinfo {pages} {041039} (\bibinfo {year} {2016})}\BibitemShut
  {NoStop}%
\bibitem [{\citenamefont {Larsen}(1963)}]{larsen1963}%
  \BibitemOpen
  \bibfield  {author} {\bibinfo {author} {\bibfnamefont {D.~M.}\ \bibnamefont
  {Larsen}},\ }\href@noop {} {\bibfield  {journal} {\bibinfo  {journal} {Annals
  of Physics}\ }\textbf {\bibinfo {volume} {24}},\ \bibinfo {pages} {89}
  (\bibinfo {year} {1963})}\BibitemShut {NoStop}%
\bibitem [{Note1()}]{Note1}%
  \BibitemOpen
  \bibinfo {note} {A gradient expansion for a one-component BEC yields higher
  order derivative terms~\cite {Braaten1997}. They will however be subleading
  compared to the kinetic term in Eq.~\protect \textup {\hbox {\mathsurround
  \z@ \protect \normalfont (\ignorespaces \ref {eq:LHYEnergy}\unskip
  \@@italiccorr )}}.}\BibitemShut {Stop}%
\bibitem [{SM()}]{SM}%
  \BibitemOpen
  \href@noop {} {}\bibinfo {note} {See Supplemental Material for
  details.}\BibitemShut {Stop}%
\bibitem [{\citenamefont {Lysebo}\ and\ \citenamefont
  {Veseth}(2010)}]{lysebo2010}%
  \BibitemOpen
  \bibfield  {author} {\bibinfo {author} {\bibfnamefont {M.}~\bibnamefont
  {Lysebo}}\ and\ \bibinfo {author} {\bibfnamefont {L.}~\bibnamefont
  {Veseth}},\ }\href@noop {} {\bibfield  {journal} {\bibinfo  {journal} {Phys.
  Rev. A}\ }\textbf {\bibinfo {volume} {81}},\ \bibinfo {pages} {032702}
  (\bibinfo {year} {2010})}\BibitemShut {NoStop}%
\bibitem [{\citenamefont {Antoine}\ and\ \citenamefont
  {Duboscq}(2014)}]{antoine2014}%
  \BibitemOpen
  \bibfield  {author} {\bibinfo {author} {\bibfnamefont {X.}~\bibnamefont
  {Antoine}}\ and\ \bibinfo {author} {\bibfnamefont {R.}~\bibnamefont
  {Duboscq}},\ }\href@noop {} {\bibfield  {journal} {\bibinfo  {journal}
  {Computer Physics Communications}\ }\textbf {\bibinfo {volume} {185}},\
  \bibinfo {pages} {2969} (\bibinfo {year} {2014})}\BibitemShut {NoStop}%
\bibitem [{\citenamefont {Antoine}\ and\ \citenamefont
  {Duboscq}(2015)}]{antoine2015}%
  \BibitemOpen
  \bibfield  {author} {\bibinfo {author} {\bibfnamefont {X.}~\bibnamefont
  {Antoine}}\ and\ \bibinfo {author} {\bibfnamefont {R.}~\bibnamefont
  {Duboscq}},\ }\href@noop {} {\bibfield  {journal} {\bibinfo  {journal}
  {Computer Physics Communications}\ }\textbf {\bibinfo {volume} {193}},\
  \bibinfo {pages} {95} (\bibinfo {year} {2015})}\BibitemShut {NoStop}%
\bibitem [{\citenamefont {Jin}\ \emph {et~al.}(1996)\citenamefont {Jin},
  \citenamefont {Ensher}, \citenamefont {Matthews}, \citenamefont {Wieman},\
  and\ \citenamefont {Cornell}}]{jin1996}%
  \BibitemOpen
  \bibfield  {author} {\bibinfo {author} {\bibfnamefont {D.}~\bibnamefont
  {Jin}}, \bibinfo {author} {\bibfnamefont {J.}~\bibnamefont {Ensher}},
  \bibinfo {author} {\bibfnamefont {M.}~\bibnamefont {Matthews}}, \bibinfo
  {author} {\bibfnamefont {C.}~\bibnamefont {Wieman}}, \ and\ \bibinfo {author}
  {\bibfnamefont {E.}~\bibnamefont {Cornell}},\ }\href@noop {} {\bibfield
  {journal} {\bibinfo  {journal} {Phys. Rev. Lett.}\ }\textbf {\bibinfo
  {volume} {77}},\ \bibinfo {pages} {420} (\bibinfo {year} {1996})}\BibitemShut
  {NoStop}%
\bibitem [{\citenamefont {Stamper-Kurn}\ \emph {et~al.}(1998)\citenamefont
  {Stamper-Kurn}, \citenamefont {Miesner}, \citenamefont {Inouye},
  \citenamefont {Andrews},\ and\ \citenamefont {Ketterle}}]{stamper1998}%
  \BibitemOpen
  \bibfield  {author} {\bibinfo {author} {\bibfnamefont {D.}~\bibnamefont
  {Stamper-Kurn}}, \bibinfo {author} {\bibfnamefont {H.-J.}\ \bibnamefont
  {Miesner}}, \bibinfo {author} {\bibfnamefont {S.}~\bibnamefont {Inouye}},
  \bibinfo {author} {\bibfnamefont {M.}~\bibnamefont {Andrews}}, \ and\
  \bibinfo {author} {\bibfnamefont {W.}~\bibnamefont {Ketterle}},\ }\href@noop
  {} {\bibfield  {journal} {\bibinfo  {journal} {Phys. Rev. Lett.}\ }\textbf
  {\bibinfo {volume} {81}},\ \bibinfo {pages} {500} (\bibinfo {year}
  {1998})}\BibitemShut {NoStop}%
\bibitem [{\citenamefont {Dalfovo}\ \emph {et~al.}(1999)\citenamefont
  {Dalfovo}, \citenamefont {Giorgini}, \citenamefont {Pitaevskii},\ and\
  \citenamefont {Stringari}}]{dalfovo1999}%
  \BibitemOpen
  \bibfield  {author} {\bibinfo {author} {\bibfnamefont {F.}~\bibnamefont
  {Dalfovo}}, \bibinfo {author} {\bibfnamefont {S.}~\bibnamefont {Giorgini}},
  \bibinfo {author} {\bibfnamefont {L.~P.}\ \bibnamefont {Pitaevskii}}, \ and\
  \bibinfo {author} {\bibfnamefont {S.}~\bibnamefont {Stringari}},\ }\href@noop
  {} {\bibfield  {journal} {\bibinfo  {journal} {Rev. Mod. Phys.}\ }\textbf
  {\bibinfo {volume} {71}},\ \bibinfo {pages} {463} (\bibinfo {year}
  {1999})}\BibitemShut {NoStop}%
\bibitem [{\citenamefont {Pethick}\ and\ \citenamefont
  {Smith}(2002)}]{pethick2002}%
  \BibitemOpen
  \bibfield  {author} {\bibinfo {author} {\bibfnamefont {C.~J.}\ \bibnamefont
  {Pethick}}\ and\ \bibinfo {author} {\bibfnamefont {H.}~\bibnamefont
  {Smith}},\ }\href@noop {} {\emph {\bibinfo {title} {Bose-Einstein
  condensation in dilute gases}}}\ (\bibinfo  {publisher} {Cambridge university
  press},\ \bibinfo {year} {2002})\BibitemShut {NoStop}%
\bibitem [{\citenamefont {Riedl}\ \emph {et~al.}(2008)\citenamefont {Riedl},
  \citenamefont {S\'anchez~Guajardo}, \citenamefont {Kohstall}, \citenamefont
  {Altmeyer}, \citenamefont {Wright}, \citenamefont {Denschlag}, \citenamefont
  {Grimm}, \citenamefont {Bruun},\ and\ \citenamefont {Smith}}]{Riedl2008}%
  \BibitemOpen
  \bibfield  {author} {\bibinfo {author} {\bibfnamefont {S.}~\bibnamefont
  {Riedl}}, \bibinfo {author} {\bibfnamefont {E.~R.}\ \bibnamefont
  {S\'anchez~Guajardo}}, \bibinfo {author} {\bibfnamefont {C.}~\bibnamefont
  {Kohstall}}, \bibinfo {author} {\bibfnamefont {A.}~\bibnamefont {Altmeyer}},
  \bibinfo {author} {\bibfnamefont {M.~J.}\ \bibnamefont {Wright}}, \bibinfo
  {author} {\bibfnamefont {J.~H.}\ \bibnamefont {Denschlag}}, \bibinfo {author}
  {\bibfnamefont {R.}~\bibnamefont {Grimm}}, \bibinfo {author} {\bibfnamefont
  {G.~M.}\ \bibnamefont {Bruun}}, \ and\ \bibinfo {author} {\bibfnamefont
  {H.}~\bibnamefont {Smith}},\ }\href@noop {} {\bibfield  {journal} {\bibinfo
  {journal} {Phys. Rev. A}\ }\textbf {\bibinfo {volume} {78}},\ \bibinfo
  {pages} {053609} (\bibinfo {year} {2008})}\BibitemShut {NoStop}%
\bibitem [{\citenamefont {Harber}\ \emph {et~al.}(2002)\citenamefont {Harber},
  \citenamefont {Lewandowski}, \citenamefont {McGuirk},\ and\ \citenamefont
  {Cornell}}]{harber2002}%
  \BibitemOpen
  \bibfield  {author} {\bibinfo {author} {\bibfnamefont {D.~M.}\ \bibnamefont
  {Harber}}, \bibinfo {author} {\bibfnamefont {H.~J.}\ \bibnamefont
  {Lewandowski}}, \bibinfo {author} {\bibfnamefont {J.~M.}\ \bibnamefont
  {McGuirk}}, \ and\ \bibinfo {author} {\bibfnamefont {E.~A.}\ \bibnamefont
  {Cornell}},\ }\href@noop {} {\bibfield  {journal} {\bibinfo  {journal} {Phys.
  Rev. A}\ }\textbf {\bibinfo {volume} {66}},\ \bibinfo {pages} {053616}
  (\bibinfo {year} {2002})}\BibitemShut {NoStop}%
\bibitem [{\citenamefont {Wild}\ \emph {et~al.}(2012)\citenamefont {Wild},
  \citenamefont {Makotyn}, \citenamefont {Pino}, \citenamefont {Cornell},\ and\
  \citenamefont {Jin}}]{wild2012}%
  \BibitemOpen
  \bibfield  {author} {\bibinfo {author} {\bibfnamefont {R.~J.}\ \bibnamefont
  {Wild}}, \bibinfo {author} {\bibfnamefont {P.}~\bibnamefont {Makotyn}},
  \bibinfo {author} {\bibfnamefont {J.~M.}\ \bibnamefont {Pino}}, \bibinfo
  {author} {\bibfnamefont {E.~A.}\ \bibnamefont {Cornell}}, \ and\ \bibinfo
  {author} {\bibfnamefont {D.~S.}\ \bibnamefont {Jin}},\ }\href@noop {}
  {\bibfield  {journal} {\bibinfo  {journal} {Phys. Rev. Lett.}\ }\textbf
  {\bibinfo {volume} {108}},\ \bibinfo {pages} {145305} (\bibinfo {year}
  {2012})}\BibitemShut {NoStop}%
\bibitem [{\citenamefont {J\o{}rgensen}\ \emph {et~al.}(2016)\citenamefont
  {J\o{}rgensen}, \citenamefont {Wacker}, \citenamefont {Skalmstang},
  \citenamefont {Parish}, \citenamefont {Levinsen}, \citenamefont
  {Christensen}, \citenamefont {Bruun},\ and\ \citenamefont
  {Arlt}}]{jorgensen2016}%
  \BibitemOpen
  \bibfield  {author} {\bibinfo {author} {\bibfnamefont {N.~B.}\ \bibnamefont
  {J\o{}rgensen}}, \bibinfo {author} {\bibfnamefont {L.}~\bibnamefont
  {Wacker}}, \bibinfo {author} {\bibfnamefont {K.~T.}\ \bibnamefont
  {Skalmstang}}, \bibinfo {author} {\bibfnamefont {M.~M.}\ \bibnamefont
  {Parish}}, \bibinfo {author} {\bibfnamefont {J.}~\bibnamefont {Levinsen}},
  \bibinfo {author} {\bibfnamefont {R.~S.}\ \bibnamefont {Christensen}},
  \bibinfo {author} {\bibfnamefont {G.~M.}\ \bibnamefont {Bruun}}, \ and\
  \bibinfo {author} {\bibfnamefont {J.~J.}\ \bibnamefont {Arlt}},\ }\href@noop
  {} {\bibfield  {journal} {\bibinfo  {journal} {Phys. Rev. Lett.}\ }\textbf
  {\bibinfo {volume} {117}},\ \bibinfo {pages} {055302} (\bibinfo {year}
  {2016})}\BibitemShut {NoStop}%
\bibitem [{\citenamefont {Fetter}\ and\ \citenamefont
  {Walecka}(1971)}]{Fetter1971}%
  \BibitemOpen
  \bibfield  {author} {\bibinfo {author} {\bibfnamefont {A.}~\bibnamefont
  {Fetter}}\ and\ \bibinfo {author} {\bibfnamefont {J.}~\bibnamefont
  {Walecka}},\ }\href@noop {} {\emph {\bibinfo {title} {Quantum Theory of
  Many-Particle Systems}}},\ Dover Books on Physics Series\ (\bibinfo
  {publisher} {Dover Publications},\ \bibinfo {year} {1971})\BibitemShut
  {NoStop}%
\bibitem [{\citenamefont {Christensen}\ \emph {et~al.}(2015)\citenamefont
  {Christensen}, \citenamefont {Levinsen},\ and\ \citenamefont
  {Bruun}}]{christensen2015}%
  \BibitemOpen
  \bibfield  {author} {\bibinfo {author} {\bibfnamefont {R.~S.}\ \bibnamefont
  {Christensen}}, \bibinfo {author} {\bibfnamefont {J.}~\bibnamefont
  {Levinsen}}, \ and\ \bibinfo {author} {\bibfnamefont {G.~M.}\ \bibnamefont
  {Bruun}},\ }\href@noop {} {\bibfield  {journal} {\bibinfo  {journal} {Phys.
  Rev. Lett.}\ }\textbf {\bibinfo {volume} {115}},\ \bibinfo {pages} {160401}
  (\bibinfo {year} {2015})}\BibitemShut {NoStop}%
\bibitem [{\citenamefont {Utesov}\ \emph {et~al.}(2018)\citenamefont {Utesov},
  \citenamefont {Baglay},\ and\ \citenamefont {Andreev}}]{Utesov2018}%
  \BibitemOpen
  \bibfield  {author} {\bibinfo {author} {\bibfnamefont {O.~I.}\ \bibnamefont
  {Utesov}}, \bibinfo {author} {\bibfnamefont {M.~I.}\ \bibnamefont {Baglay}},
  \ and\ \bibinfo {author} {\bibfnamefont {S.~V.}\ \bibnamefont {Andreev}},\
  }\href@noop {} {\bibfield  {journal} {\bibinfo  {journal} {Phys. Rev. A}\
  }\textbf {\bibinfo {volume} {97}},\ \bibinfo {pages} {053617} (\bibinfo
  {year} {2018})}\BibitemShut {NoStop}%
\bibitem [{\citenamefont {Braaten}\ and\ \citenamefont
  {Nieto}(1997)}]{Braaten1997}%
  \BibitemOpen
  \bibfield  {author} {\bibinfo {author} {\bibfnamefont {E.}~\bibnamefont
  {Braaten}}\ and\ \bibinfo {author} {\bibfnamefont {A.}~\bibnamefont
  {Nieto}},\ }\href@noop {} {\bibfield  {journal} {\bibinfo  {journal} {Phys.
  Rev. B}\ }\textbf {\bibinfo {volume} {56}},\ \bibinfo {pages} {14745}
  (\bibinfo {year} {1997})}\BibitemShut {NoStop}%
\end{thebibliography}%


\clearpage

\setcounter{figure}{0}
\makeatletter 
\renewcommand{\thefigure}{S\@arabic\c@figure}
\makeatother

\onecolumngrid

\subsection*{\large Supplemental Material to ``Dilute Fluid Governed by Quantum Fluctuations''}

In this supplemental material, we show additional equations and numerical calculations, which complement the results shown in the main text.

In a Bose-Bose mixture of equal masses $m$, the contribution from quantum fluctuations to the local energy density is~\cite{larsen1963, petrov2015}
\begin{align}\tag{S1}
\frac{E_\text{LHY}}{V} & = \frac{32\sqrt{2\pi}}{15} \frac{\hbar^2}{m} \sum_{\pm} \left( a_{11}n_1 + a_{22}n_2 \pm \sqrt{(a_{11}n_1 - a_{22}n_2)^2 + 4a_{12}^2n_1n_2} \right)^{5/2},
\label{eq:E_LHY_1}
\end{align}
where $n_1$ and $n_2$ are the densities of the respective components. The interaction strengths within the two components are parameterized by the scattering lengths $a_{11}$ and $a_{22}$, and the interaction strength between the components is described by $a_{12}$. For the case of $a_{12}=-\sqrt{a_{11}a_{22}}$, Eq.~(\ref{eq:E_LHY_1}) reduces to
\begin{align}\tag{S2}
\frac{E_\text{LHY}}{V} = \frac{256\sqrt{\pi}}{15} \frac{\hbar^2}{m} \left( a_{11} n_1 + a_{22} n_2 \right)^{5/2},
\label{eq:E_LHY_2}
\end{align}
and for $n_1/n_2 = \sqrt{a_{22}/{a_{11}}}$, it further reduces to 
\begin{align}\tag{S3}
\frac{E_\text{LHY}}{V} = \frac{256\sqrt{\pi}}{15} \frac{\hbar^2}{m} \left( n |a_{12}| \right)^{5/2},
\label{eq:E_LHY_3}
\end{align}
where $n = n_1+n_2$, and the two-component framework is reduced to a one-component framework.

To derive a Gross-Piteavskii equation for the Bose-Bose mixture, it is necessary to evaluate the contribution of quantum fluctuations to the chemical potential of each component $i$
\begin{align}\tag{S4}
\mu_\text{LHY}^{(i)} = \frac{\partial E_\text{LHY}}{\partial N_i} = \frac{\partial (E_\text{LHY}/V)}{\partial n_i},
\end{align}
where $N_i$ refers to the particle number. 

Based on Eq.~(\ref{eq:E_LHY_1}), one obtains
\begin{align}\tag{S5}
\mu_\text{LHY}^{(i)} = \frac{16\sqrt{2\pi}}{3} \frac{\hbar^2}{m} \sum_\pm \Bigg[ 
& \left( a_{ii} \pm \frac{a_{ii}^2n_i - a_{ii} a_{jj} n_j + 2 a_{ij}^2 n_j}{\sqrt{(a_{ii}n_i - a_{jj}n_j)^2 + 4a_{ij}n_i n_j}} \right) \nonumber \\
\times & \left( a_{ii} n_i + a_{jj} n_j \pm \sqrt{(a_{ii}n_i - a_{jj}n_j)^2 + 4a_{ij}n_i n_j} \right)^{3/2} \Bigg].
\end{align}
This expression is used for the results shown in Fig.~4 in the main text.

When $a_{12}=-\sqrt{a_{11}a_{22}}$, as for Eq.~(\ref{eq:E_LHY_2}), the addition to the chemical potential is
\begin{align}\tag{S6}
\mu_\text{LHY}^{(i)} = \frac{128\sqrt{\pi}}{3} \frac{\hbar^2}{m} a_{ii} \left( a_{11} n_{1} + a_{22} n_{2} \right)^{3/2} .
\label{eq:mu_LHY_2}
\end{align}
This expression is used for the results for the two-component calculations shown in Figs.~1-3 of the main text.

For the one-component framework where $n_1/n_2 = \sqrt{a_{22}/{a_{11}}}$, the addition to the chemical potentials from Eq.~(\ref{eq:E_LHY_3}) is
\begin{align}\tag{S7}
\mu_\text{LHY} =  \frac{\partial (E_\text{LHY}/V)}{\partial n} =\frac{128\sqrt{\pi}}{3} \frac{\hbar^2}{m} n^{3/2} |a_{12}|^{5/2} ,
\end{align}
which is used for Eq.~(5), and the one-component results shown in Figs.~1-3 of the main text.

The coupled two-component Gross-Piteavskii equations based on the result of Eq.~(\ref{eq:mu_LHY_2}) are
\begin{align}\tag{S8}
\mu_i \Psi_i= \left[ - \frac{\hbar^2}{2m} \nabla^2 + V + \frac{128\sqrt{\pi}}{3} \frac{\hbar^2}{m} a_{ii} \left( a_{11} |\Psi_1|^2 + a_{22}  |\Psi_2|^2 \right)^{3/2} \right] \Psi_i,
\end{align}
where $V$ is the external potential and $\Psi_i$ is the condensate wave function. We use $\mu_i$ calculated from the expectation value of the right hand side of this equation to derive Eq.~(16) of the main text.

In the following, we present numerical calculations which extend the results shown in the main text towards stronger interactions.

\begin{figure}[h]
	\centering
	\includegraphics{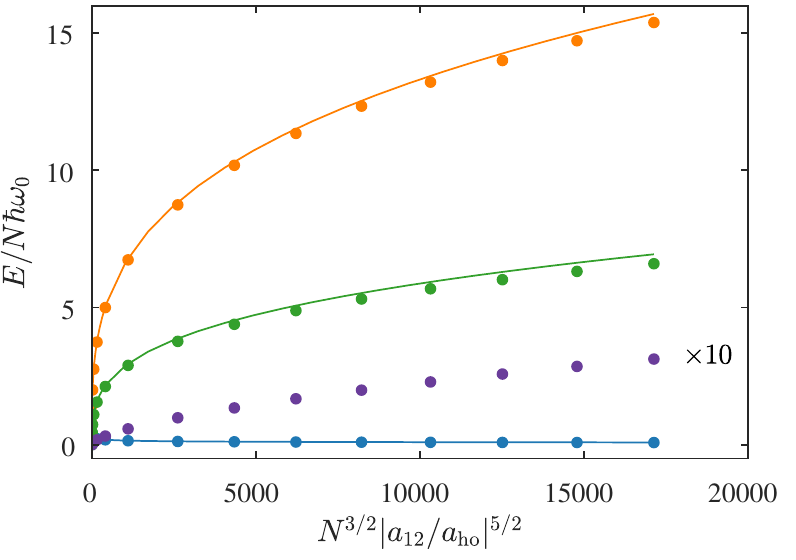}
	\caption{Numerically obtained energies of the LHY fluid, similar to Fig.~1 of the main text. The lines refer to one-component calculations, and the points to two-component calculations. From top to bottom: Orange is potential energy, green is the LHY correction, purple is the mean-field energy, and blue is kinetic energy. Mean-field energy only exists in the two-component framework and has been multiplied by a factor $10$ for visibility.}
\label{figS:energies}
\end{figure}

Figure~\ref{figS:energies} shows ground-state energies of the LHY fluid calculated numerically using the one-component and the two-component framework. It is similar to Fig.~1 of the main text, but extended to stronger interactions. The numerical calculations were performed using $N=10^5$, $a_{12} = 53 a_0$, and trap frequencies up to $\omega_0/2\pi = \SI{80}{kHz}$. A trap frequency this large is typically not available experimentally, and the main purpose of these calculations is to show the validity of the one-component framework when approaching strong interactions. Indeed, the results display that the one-component framework approximates the two-component system well, even far beyond the experimentally available regime.

\begin{figure}[h]
	\centering
	\includegraphics{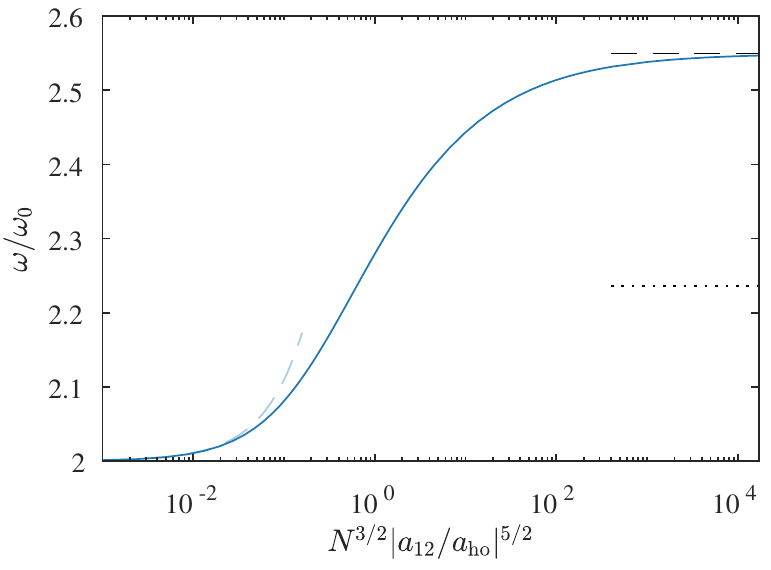}
	\caption{Monopole frequency of the LHY fluid. The light blue dashed line is Eq.~(15), valid for weak interactions. The black 
	dashed and dotted lines give the Thomas-Fermi results $\omega/ \omega_0 = \sqrt{13/2}$ and $\omega/ \omega_0 = \sqrt{5}$ for a
	 LHY fluid and a regular BEC, respectively. The full blue line is obtained from Eq.~(14) of the main text combined with ground-state calculations performed in the one-component framework.}
\label{figS:osc_freqs}
\end{figure}

Figure~\ref{figS:osc_freqs} shows the monopole frequency of the LHY fluid, using  Eq.~(14) of the main text and the one-component ground-state calculations shown in Fig.~\ref{figS:energies}. The figure is similar to Fig.~2 of the main text, but extended to stronger interactions and using a logarithmic axis. These result more clearly show how the monopole frequency approaches the Thomas-Fermi limit prediction when the interaction strength is increased.

\end{document}